\documentclass[aps,prb,showpacs,twocolumn]{revtex4}

\usepackage{graphicx,amsmath}
\usepackage{epsfig}
\usepackage{color}

\def  \bnabla  {\mbox{\boldmath$\nabla$}}
\begin{document}

\title{Nonlinear spin Hall effect in GaAs (110) quantum wells}
\author{V. I. Ivanov$^{1}$, V. K. Dugaev$^{2,3}$, E. Ya. Sherman$^{4,5}$, and J.
Barna\'s$^{6,*}$} \affiliation{$^{1}$Insitute for Problems of
Materials Science, Ukrainian
Academy of Sciences, Vilde 5, 58001 Chernovtsy, Ukraine\\
$^{2}$Department of Physics, Rzesz\'{o}w University of Technology,
al.~Powsta\'{n}c\'{o}w Warszawy 6, 35-959 Rzesz\'{o}w, Poland \\
$^{3}$Department of Physics and CFIF, Instituto Superior T\'{e}cnico, TU
Lisbon, Av.~Rovisco Pais 1049-001 Lisbon, Portugal \\
$^{4}$Department of Physical Chemistry, Universidad del Pa\'{i}s Vasco,
Bilbao, Spain \\
$^{5}$IKERBASQUE Basque Foundation for Science, 48011, Bilbao, Spain \\
$^{6}$Institute of Molecular Physics, Polish Academy of Sciences,
ul.~Smoluchowskiego 17, 60-179 Pozna\'{n}, Poland}

\begin{abstract}
We consider stationary spin current in a (110)-oriented GaAs-based
symmetric quantum well due to a nonlinear response to an external
periodic electric field. The model assumed includes the
Dresselhaus spin-orbit interaction and the random Rashba
spin-orbit coupling. The Dresselhaus term is uniform in the
quantum well plane and gives rise to spin splitting of the
electron band. The external electric field of frequency $\omega$
-- in the presence of random Rashba coupling -- leads to virtual
spin-flip transitions between spin subbands, generating stationary
pure spin current proportional to the square of the field
amplitude.
\end{abstract}
\date{\today }

\maketitle

\section{Introduction}

Spin-orbit (SO) interactions in semiconductors reveal a variety of
fundamental spin related phenomena.\cite{zutic04} Formation of
stable spin helices with nontrivial temporal and spatial dynamics,
\cite {Koralek09,Pershin10,Nunner11} spin optics,\cite{Khodas04}
and spin-dependent sound radiation \cite{Smirnov11} represent only
some of these effects. Spin currents,\cite{Sinova04} solely
attributed to the spin-orbit coupling, provide a possibility of
inducing and controlling spin motion by electrical and optical
fields, and therefore became one of the key elements of the modern
spintronics oriented at new device applications of semiconductor
based structures. Thorough investigations of realistic systems
enlarge the variety of both fundamental phenomena and possible
applications. As an example we mention that disorder -- always
present in real systems -- plays a crucial role in the
spin-Hall effect \cite{disorder}, as the spin Hall conductivity
can be totally suppressed by any finite concentration of
impurities.

Recently, symmetric GaAs (110) quantum wells (QWs) became the
subject of extensive experimental and theoretical investigations.
This is related to the expectation of the longest spin relaxation
times in these structures,
\cite{Ohno99,Dohrmann04,Hankiewicz06,Schreiber07,Belkov08,Crankshaw09,Tarasenko09,zhou09}
which in turn can lead to interesting spin
dynamics.\cite{Volkl,Hubner} A stationary pure spin current
accompanying an electric current in (110) QWs was observed as
reported in Ref.~[\onlinecite{Sih}]. SO interaction in these
systems, described by the Dresselhaus term in the corresponding
Hamiltonian, \cite{dresselhaus55,winkler04,Dyakonov86} conserves
the electron spin along the axis normal to the QW plane for any
electron momentum $\mathbf{k}$. As a result, random motion of an
electron does not lead to a random direction of the spin-orbit
field and therefore does not lead to spin relaxation. In reality,
however, this spin component relaxes very slowly, and its analysis
provides a test for the rapidly developing low-frequency
spin-noise spectroscopy \cite{Muller08} suitable for the
measurements of the long spin evolutions.

In the case of perfect $z\to -z$ symmetry (the axis $z$ is
perpendicular to the QW plane), the Rashba SO interaction is zero.
In real structures, however, the Rashba coupling still exists in
the form of a spatially fluctuating SO field (though being zero on
average). \cite{sherman03,sherman05,glazov10}
This interaction induces spin-flip processes leading to the spin relaxation,%
\cite{zhou09} and can be also responsible for generation of a
nonequilibrium spin density due to the absorption of an external
electromagnetic field. \cite{glazov10} Recently, it was proposed
that this random SO coupling can result in the spin orientation by
an external current \cite{Golub} and also can play a role in the
formation of the stripe structure of spin current
distribution.\cite{Lyanda}

In this paper we propose a new possibility of exciting a steady
pure spin current by a periodic external filed, extending thus the
abilities of spin manipulation in real situations. In contrast to
the conventional spin Hall effect, which is linear in the external
electric field, the proposed spin current is quadratic in the
external periodic field. The effect is a result of the interplay
of constant Dresselhaus and spatially random Rashba terms, and is
not related to the spin currents produced by gate manipulation of
the Rashba coupling \cite{Tang} or adiabatic pumping in
graphene.\cite{Zhang}  Exact mechanism of the effect does not
necessarily involve real spin-flip transitions of electrons between the
spin-split subbands in (110)-oriented GaAs QW, but relies on
virtual spin-flip processes which renormalize the wave functions
of electrons in a nonequilibrium state. This makes such a
nonlinear current a physically new phenomenon, which appears if
one accounts for more realistic effects than those described
by the conventional Rashba and Dresselhaus models. 

In Section 2 we
describe the model and Hamiltonian of the system. Spin current is
calculated in Section 3. Summary and final conclusions are presented in
Section 4.

\section{Model}

Hamiltonian of a two-dimensional electron gas with the constant
Dresselhaus term $H_{D}$ and spatially fluctuating Rashba
spin-orbit interaction $H_{R}$, subjected to external
electromagnetic field described by the vector
potential $\mathbf{A}(\mathbf{r},t)$, takes the following
form (we use units with $\hbar =1$)
\begin{equation}
H=H_{0}+H_{D}+H_{R},
\end{equation}
where the first two terms are
\begin{eqnarray}
&&H_{0}=-\frac{1}{2m}\left( \bnabla -\frac{ie\mathbf{A}}{c}\right) ^{2},
\label{H0} \\
&&H_{D}=-i\alpha \sigma _{z}\left( \nabla _{x}-\frac{ieA_{x}}{c}\right).
\label{HD}
\end{eqnarray}
The Dresselhaus constant $\alpha=\gamma\pi^{2}/2w^{2}$, where
$\gamma$ is the corresponding bulk Dresselhaus coupling parameter,
is inversely proportional to the square of the QW width $w$. The
other components of the Dresselhaus interaction vanish due to the
specific symmetry of the (110) orientation.\cite{winkler04,zhou09}

The last term in Eq.~(1) stands for the effects of the spatially
nonuniform Rashba SO interaction, which can be written as 
$H_R=H_R^0+V$, where $H_R^0$ is the Rashba term for $\mathbf{A}%
(\mathbf{r},t)=0$,
\begin{equation}
H_{R}^0=-\frac{i}{2}\sigma _{x}\left\{ \nabla _{y},\,\lambda (\mathbf{r}%
)\right\} +\frac{i}{2}\sigma _{y}\left\{ \nabla _{x},\,\lambda (\mathbf{r}%
)\right\},  \label{HR0}
\end{equation}
with $\left\{ \,,\right\} $ denoting the anticommutator and $\lambda (%
\mathbf{r})$ being the random Rashba SO interaction. The
term $V$, in turn, describes coupling of the electron spin to the external
field $\mathbf{A}(\mathbf{r},t)$ \textit{via} the Rashba field,
\begin{equation}
V=-\frac{e}{c}\lambda(\mathbf{r})\left( \sigma _{x}A_{y}-\sigma
_{y}A_{x}\right) .  \label{V}
\end{equation}
Due to the assumed symmetry with respect to $z$-inversion, the spatially
averaged Rashba interaction vanishes, $\left\langle \lambda (\mathbf{r}%
)\right\rangle =0.$ We assume that the random Rashba field can be described
by the correlation function related to fluctuating density of impurities
near the QW,\cite{sherman03,glazov10}
\begin{equation}
C_{\lambda \lambda }\left( \mathbf{r-r}^{\prime }\right) \equiv \langle
\lambda (\mathbf{r})\,\lambda (\mathbf{r}^{\prime })\rangle =\left\langle
\lambda ^{2}\right\rangle F\left( \mathbf{r-r}^{\prime }\right) ,  \label{7}
\end{equation}
where the range function $F\left( \mathbf{r-r}^{\prime }\right) $ depends on
the type of disorder. We assume the correlator of random Rashba interaction
in the momentum space in the form\cite{glazov10,Dugaev09}
\begin{equation}
|\lambda _{q}|^{2}=2\pi \left\langle \lambda ^{2}\right\rangle
R^{2}\,e^{-qR},
\end{equation}
where $R$ is the spatial scale of the fluctuations.

In the absence of external field and random Rashba SO interaction,
the Hamiltonian $H_{0}+H_{D}$ describes the spectrum of
spin-polarized electrons, $\varepsilon
_{\mathbf{k}\sigma}=\left(k_{x}^{2}+k_{y}^{2}\right)/2m+\sigma
\alpha k_{x}$. The energy bands of spin up and spin down electrons
are thus shifted in opposite directions along the $k_{x}$ axis.
The corresponding Green function is then diagonal in the spin
subspace,
\begin{eqnarray}
&&\mathbf{G}_{\mathbf{k}\varepsilon }^{(0)}=
\left(
\begin{array}{cc}
{G}_{{\mathbf{k}}\varepsilon +} &  0 \\
0 & {G}_{{\mathbf{k}}\varepsilon -}
\end{array}
\right),  
\nonumber \\
&& {G}_{\mathbf{k}\varepsilon\sigma}= \frac{1}
{\varepsilon-\varepsilon_{\mathbf{k}\sigma}+\mu 
+i\delta _{\mathbf{k}\sigma}\, \mathrm{sign}({\varepsilon})}, \label{Green}
\end{eqnarray}
where $\sigma=+$ for spin up ($\uparrow$) electrons and $\sigma=-
$ for spin down ($\downarrow$) electrons, whereas $\delta
_{\mathbf{k}\sigma}$ is the momentum and spin dependent relaxation
rate.

\section{Nonlinear second-order spin current}

In the following we consider the $z$-component of a pure spin
current flowing along the $x$ axis, that is the only component allowed
by symmetry of the system under consideration. The operators of
the electron velocity $\hat{v}_{x}$ and the corresponding spin
current tensor component $\hat{\jmath}_{x}^{z}$ are
\begin{eqnarray}
&&\hat{v}_{x}=i[H_{0}+H_{D},x]=\frac{k_{x}}{m}+\alpha \sigma _{z}-\lambda
\sigma _{y},  \label{5} \\
&&\hat{\jmath}_{x}^{z}=\frac{1}{2}\left\{ \hat{v}_{x},\sigma _{z}\right\} =%
\frac{k_{x}}{m}\,\sigma _{z}+\alpha,  \label{spincurrent}
\end{eqnarray}
where the $\alpha$-related terms correspond to the anomalous contribution
to the velocity. The macroscopic spin current density is then given by
\begin{equation}
j_{x}^{z}=i\,\mathrm{Tr}\,\sum_{\mathbf{k}}\int \frac{d\varepsilon }{2\pi }\,%
\hat{\jmath}_{x}^{z}\,\mathbf{G}_{\mathbf{k}\varepsilon },  \label{8}
\end{equation}
where $\mathbf{G}_{\mathbf{k}\varepsilon }$ is the Green's
function of the system interacting with the external
electromagnetic field. 

Upon substituting (10) into Eq.~(11) one
can note that the second term describes the current caused only by
the electron density,
\begin{equation}
n=i\,\mathrm{Tr}\sum_{\mathbf{k}}
\int\frac{d\varepsilon}{2\pi}\,\mathbf{G}_{\mathbf{k}\varepsilon
},
\end{equation}
conserved under any external perturbation. This conservation is
achieved in calculations by an appropriate shift of the chemical
potential $\mu $.
In equilibrium, however, there is no spin current in the system
(expected, e.g., for QWs with other crystallographic orientations)
since the integrated contributions from $k_{x}/m$ and $\alpha$
terms in Eq.~(\ref{spincurrent}) exactly cancel each other. As a
result, this type of structures does not demonstrate the Rashba
paradox of the non-zero equilibrium pure spin
current.\cite{Rashba} This can be seen directly by calculating
spin current using Eq.~(\ref{8}) with the equilibrium Green
function in Eq.~(\ref{Green}), or by taking into account the fact
that Hamiltonian in Eq.~(\ref{HD}) can be transformed by the
$SU(2)$ rotation to the form that does not have spin dependent
terms.\cite{Tokatly}

\begin{figure}[tbp]
\center
\includegraphics[scale=0.45]{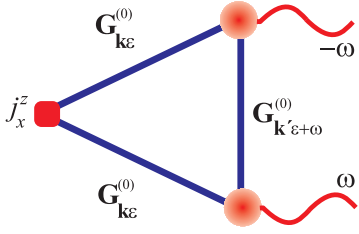}
\caption{Feynman diagrams for the excited spin current in
Eq.~(\ref{8}). The  circles correspond to the matrix elements
$V_{\mathbf{kk^{\prime}}}$ and $V_{\mathbf{k^{\prime}k}}$. }
\label{diagrams}
\end{figure}

However, a nonzero pure spin current, which is the subject of 
interest here, can be generated by an external field in the
presence of random Rashba coupling, as presented schematically by
the Feynman graph in Fig.~\ref{diagrams}. In this graph we
introduced the following notations:
\begin{eqnarray}\label{Vkk}
&&V_{\mathbf{kk^{\prime }}}=\lambda _{\mathbf{kk^{\prime }}}\left( \sigma
_{x}A_{y}(\omega )-\sigma _{y}A_{x}(\omega )\right) , \\
&&V_{\mathbf{k^{\prime }k}}=\lambda _{\mathbf{k^{\prime }k}}\left(
\sigma _{x}A_{y}(-\omega )-\sigma _{y}A_{x}(-\omega )\right) ,
\end{eqnarray}
for the transition matrix elements due to the external field and spin-orbit
coupling.

With the Feynman graph shown in Fig.~\ref{diagrams} we find the
spin current density omitting the $\alpha $ term in the velocity
operator since its contribution is conserved if no additional
electrons are injected to the system. The resulting expression for
the contribution of a single spin component to the total spin
current becomes
\begin{equation}
j_{x\sigma }=\frac{iA^{2}}{m}\sum_{\mathbf{kk^{\prime }}}\int
\frac{d\varepsilon }{2\pi }\,k_{x}\,|\lambda _{\mathbf{kk^{\prime }}}|^{2}
G_{\mathbf{k}\varepsilon \sigma }\,G_{\mathbf{k}^{\prime }\varepsilon +\omega
\sigma ^{\prime }}\,G_{\mathbf{k}\varepsilon \sigma },  \label{80}
\end{equation}
where $A$ is the vector potential amplitude, making the injected
current independent on the external field orientation. The total
spin current is given by
$j_{x}^{z}=j_{x,\sigma=1}-j_{x,\sigma=-1}$ with
$j_{x,\sigma=-1}=-j_{x,\sigma=1}$.

Using Eqs. (8) and (15), after rather tedious integration of the
product of the three Greens functions in the complex $\varepsilon$
plane, one obtains
\begin{widetext}
\begin{eqnarray}
j_{x\sigma }
&=&-\frac{A^{2}}{m}\sum_{\mathbf{kk^{\prime }}}k_{x}\,|\lambda_{\mathbf{kk^{\prime}}}|^{2}
\left[-\frac{f(\varepsilon_{\mathbf{k}\sigma })}
{\left(\varepsilon _{\mathbf{k}\sigma}-\varepsilon _{\mathbf{k}^{\prime }\sigma
^{\prime }}+\omega +i\delta _{\mathbf{k}\sigma }+i\delta _{\mathbf{k}^{\prime}\sigma ^{\prime }}\,
\mathrm{sign}\,(\varepsilon _{\mathbf{k}\sigma }+\omega -\mu )\right)^{2}}\right.
\nonumber \\
&&\left. +\frac{f(\varepsilon _{k^{\prime }\sigma ^{\prime }})}
{\left(-\varepsilon _{\mathbf{k}\sigma }+\varepsilon _{\mathbf{k}^{\prime }\sigma
^{\prime }}-\omega +i\delta _{\mathbf{k}^{\prime}\sigma ^{\prime }}+i\delta _{\mathbf{k}\sigma }\,\mathrm{%
sign}\,(\varepsilon _{\mathbf{k}^{\prime }\sigma ^{\prime}}-\omega-\mu)\right)^{2}}\right]  \nonumber\\
&=&\frac{A^{2}}{m}
\sum_{\mathbf{kq}}|\lambda _{q}|^{2}\left[
\frac{k_{x}\,[f(\varepsilon _{\mathbf{k}\sigma })-f(\varepsilon _{\mathbf{k}\sigma}+\omega )]}
{\left(\varepsilon _{\mathbf{k}\sigma }-\varepsilon _{\mathbf{k-q}\sigma ^{\prime }}
+\omega +i\delta _{+}\right)^{2}}-
\frac{(k_{x}+q_{x})\,f(\varepsilon _{\mathbf{k}\sigma ^{\prime }})}
{\left(\varepsilon _{\mathbf{k+q}\sigma}-\varepsilon _{\mathbf{k}\sigma ^{\prime }}
+\omega +i\delta _{-}\right)^{2}
}\right.  \nonumber\\
&&\left. +\frac{k_{x}\,f(\varepsilon_{\mathbf{k}\sigma }+\omega)}
{\left(\varepsilon _{\mathbf{k}\sigma }-\varepsilon _{\mathbf{k-q}\sigma ^{\prime}}
+\omega +i\delta _{-}\right)^{2}}
\right] .
\label{jxsigma}
\end{eqnarray}
\end{widetext}
Here we introduced the notation: $\delta _{+}\equiv \delta
_{\mathbf{k}\sigma }+\delta _{\mathbf{k}^{\prime}\sigma ^{\prime
}}$ and $\delta _{-}\equiv \delta _{\mathbf{k}\sigma }-\delta
_{\mathbf{k}^{\prime}\sigma ^{\prime }}$. Equation (16) shows that
the injection of spin current is a coherent effect arising due to
the change in electron wave function under the resonant
electromagnetic radiation rather than the injection due to the
two-photon absorption typical in nonlinear semiconductor optics.

Since the single electron energy $\varepsilon_{\mathbf{k}\sigma }$
can be written as 
$[{(k_{x}+\sigma\alpha m)^{2}+k_{y}^{2}}]/{2m}-{m\alpha^{2}}/{2}$, 
one can shift the
chemical potential, $\mu \to \mu +m\alpha ^{2}/2$. Then one can
write $j_{x\sigma }$ in the form,
\begin{widetext}
\begin{eqnarray}
j_{x\sigma } &=&\frac{A^{2}}{m}\sum_{\mathbf{kq}}|\lambda
_{q}|^{2}
\left[ (k_{x}-\sigma \alpha m)\,\frac{f(\varepsilon _{\mathbf{k}})
-f(\varepsilon _{\mathbf{k}}+\omega )}{\left(\mathbf{k}\cdot \mathbf{q}/m-q^{2}/2m+2\sigma \alpha
(k_{x}-q_{x})-2m\alpha ^{2}+\omega +i\delta _{+}\right)^{2}}\right.   \nonumber
\label{95} \\
&&\left. +(k_{x}-\sigma \alpha m)\,\frac{f(\varepsilon _{\mathbf{k}}+\omega )}{\left(\mathbf{k}%
\cdot \mathbf{q}/m-q^{2}/2m+2\alpha \sigma (k_{x}-q_{x})-2m\alpha
^{2}+\omega -i\sigma \delta _{-}\right)^{2}}\right.   \nonumber \\
&&\left. -(k_{x}+q_{x}+\sigma \alpha m)\,\frac{f(\varepsilon _{\mathbf{k}}+\omega )}
{\left(\mathbf{k}\cdot \mathbf{q}/m+q^{2}/2m+2\alpha \sigma (k_{x}+q_{x})+2m\alpha
^{2}+\omega -i\sigma \delta _{-}\right)^{2}}\right] .
\end{eqnarray}
\end{widetext}

\begin{figure}[tbp]
\center
\includegraphics[scale=0.7]{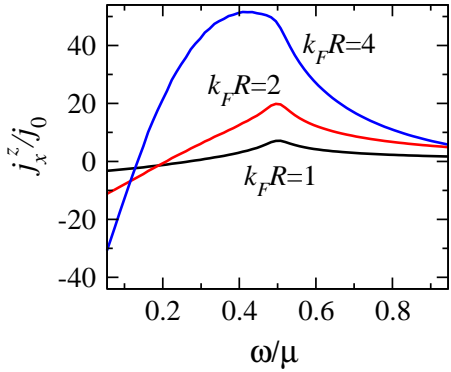}
\caption{Total spin current, calculated by using Eq.~(\ref{95})
for indicated values of $k_{F}R$. The parameters used in numerical
calculations are given in the main text.}
\label{calculated:current}
\end{figure}

We emphasize here that the calculated spin current is a stationary
coherent nonlinear effect proportional to the intensity of
incident radiation,\cite{Pershin08} in contrast to the spin
current generated by pulse excitations, where the result is
proportional to the total fluence in the pulse.\cite{sherman05}
The transition processes produce real holes in the initially
occupied subbands and electrons in those initially empty, changing
the real occupation of the spin-up and spin-down states. The
calculated current is also not related to the Drude-like linear
response at frequency $\omega $, suppressed by the factor of the
order of $\left(\delta_{+}/\omega\right)^2$.

The results of numerical calculation of the spin current (taking
part of Eq.~(\ref{95})) are presented in
Fig.~\ref{calculated:current} for different values of the parameter
$k_FR$. The parameters typical for the (110) quantum wells are:
$2\alpha m/k_{F}=0.1$ and $\delta _{+}/\mu =0.1$. For the
momentum-dependent $\delta_{-}$, we assume a typical value,
$\delta _{-}/\mu =0.05$. Furthermore, we used for GaAs:
$m=0.067\,m_{0}$, where $m_{0}$ is the free electron mass, Fermi
momentum $k_{F}=1.8\times 10^{6}$~cm$^{-1}$ (corresponding to
electron concentration $5.2\times 10^{11}$~cm$^{-2}$) and $\mu
=18.5$~meV. The spin current in Fig.~2 is presented in the units of
$j_{0}$, with $j_{0}$ defined as
\begin{equation}
j_{0}=\frac{2m^{2}\alpha e^{2}}{c^{2}\pi ^{3}}{A}^{2}
\frac{\left\langle \lambda ^{2}\right\rangle }{k_{F}^{2}}.
\end{equation}
Taking into account the relation ${A}^{2}=\left( c/\omega
\right)^{2}E^{2},$ where $E$ is the electric field amplitude, we
obtain
$j_{0}=2m^{2}\alpha\langle\lambda^{2}\rangle\left(eE/\omega\right)^{2}/\pi^{3}k_{F}^{2},$
with $eE/\omega $ being the amplitude of the momentum oscillation
of a classical electron in a periodic electric field. It is
interesting to mention that the maximum value of $E,$ which still
can be considered as a perturbation, is determined by $eE/\omega
\sim k_{F},$ and, therefore, the maximum of $j_{0}$ is of the
order of $m^{2}\alpha \left\langle \lambda ^{2}\right\rangle,$
having the physical meaning of the equilibrium spin current
induced by the random Rashba spin-orbit coupling.

\begin{figure}[tbp]
\center
\includegraphics[scale=0.6]{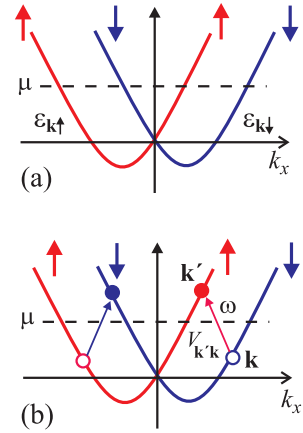}
\caption{Schematic presentation of the light-induced resonant formation of
spin holes in the energy bands occupied with electrons: (a) spin-split
energy bands in GaAs (110) quantum well; (b) due to the coupling $V_{\mathbf{%
kk^{\prime}}}$ (cf. Eq.(\ref{Vkk})) of the spin-up and spin-down
states, the effective spin $\left< S_z\right> $ in each subband
decreases.} \label{bands}
\end{figure}

To understand better the physical mechanism of the nonlinear
spin-current generation we consider a schematic picture presenting
the electron energy bands as a function of $k_{x}$ without Rashba
random SO interaction and without external field, see
Fig.~\ref{bands}. As we have already mentioned above, the
Dresselhaus SO interaction leads to spin splitting of the electron
states of a free electron gas, which results in the energy bands $%
\varepsilon _{\mathbf{k}\sigma }$ shown in Fig.~\ref{bands}(a) as
a function of $k_{x}$ (for $k_{y}=0$). Even though the states
$\left| \mathbf{k}\sigma \right\rangle $ of these bands are spin
polarized, the spin current in equilibrium is exactly zero. This
is related to the zero current associated with each of the
subbands, $j_{\uparrow ,\downarrow }$, calculated as the flux of
electrons in each subband. Obviously, vanishing current $j_{\sigma
}$ in the subband $\sigma $ means that the spin current $j_{\sigma
}^{s}$ is also zero. Distortions of the energy subbands either due
to the random Rashba interaction in Eq.~(\ref{HR0}) or due to the
external field in Eq.~(\ref {H0}) do not break the condition
$j=0$.

Our calculations, however, showed that nonzero matrix elements of
field-induced spin-flip intersubband transitions appear in the presence of
random Rashba coupling. Accordingly, in the nonequilibrium situation the
electron states in each subband are a superposition of spin up and down
states, so that the resulting state $\left| \mathbf{k}\pm \right\rangle $
(Fig.~\ref{bands}(b)) has a smaller effective spin. Such mixing of $\left|
\mathbf{k}\sigma \right\rangle $ and $\left| \mathbf{k}^{\prime }\sigma
^{\prime }\right\rangle $ states effectively depends on $|\mathbf{k}-\mathbf{%
k}^{\prime }|$ and on $|\varepsilon _{\mathbf{k}\sigma }-\varepsilon _{%
\mathbf{k}^{\prime }\sigma ^{\prime }}\pm \omega |$, so that the
above-mentioned spin mixing is different at different parts of the
dispersion curves $\varepsilon _{\mathbf{k}\sigma }$. This is shown
schematically in Fig.~\ref{bands}(b) for different spin subbands.

Thus, even
though in nonequilibrium the current in each subband $\varepsilon _{\mathbf{k%
}\sigma }$ is zero, the associated spin current is not zero
anymore. For example, in the $\varepsilon _{\mathbf{k}\uparrow }$
band more up spins
flow in $+x$ than $-x$ direction. Correspondingly, in the $\varepsilon _{%
\mathbf{k}\downarrow }$ band more down spins flow in $-x$
direction. This results in the net spin-up current in $+x$
direction. Obviously, the direction of spin current is related to
the sign of Dresselhaus coupling constant $\alpha$. A remarkable
change of the wave function by a strong electric field causes the
injected pure spin current of the order of the equilibrium spin
currents arising as a result of the Rashba
paradox.\cite{sherman05,Rashba}

\section{Summary and conclusions}

To conclude, we have proposed a new effect of  the coherent
nonlinear generation of a steady pure spin currents in GaAs (110)
quantum wells by electromagnetic wave. The injected spin current
is proportional to the intensity of the external 
radiation, strongly depends on the frequency, and can be injected
in the frequency range up to the Fermi energy of the
two-dimensional electron gas. Physical mechanism of the effect is
related to the virtual spin reorientation of electrons filling the
spin subbands split by the Dresselhaus interaction in the presence
of a randomly varying Rashba coupling. The latter may be
introduced, e.g. by random doping of the quantum well. As a
result, a 'spin hole' virtually appears in the subband, leading to
the light-induced spin current.

\textit{Acknowledgements.} This work is partly supported by FCT Grant
PTDC/FIS/70843/2006 in Portugal and by National Science Center in Poland
as a research project in years 2011 -- 2014. The work of EYS was
supported by the MCINN of Spain grant FIS2009-12773-C02-01 and "Grupos Consolidados UPV/EHU del
Gobierno Vasco" grant IT-472-10.


\begin{thebibliography}{99}

\item[*] Permanent position: Faculty of Physics, Adam Mickiewicz
University, ul.~Umultowska 85, 61-614 Pozna\'{n}, Poland.

\bibitem{zutic04}  I. Zuti\'{c}, J. Fabian, and S. Das Sarma, \rmp \textbf{76%
}, 323 (2004); M. W. Wu, J. H. Jiang, and M. Q. Weng, Phys. Reports, \textbf{%
493}, 61 (2010).

\bibitem{Koralek09}  J. D. Koralek, C. Weber, J. Orenstein, A. Bernevig,
S.-C. Zhang, S. Mack, and D. Awschalom, Nature \textbf{458}, 610 (2009).

\bibitem{Pershin10}  Yu. V. Pershin and V. A. Slipko, 
                     Phys. Rev. B \textbf{82}, 125325 (2010), 
                     V. A. Slipko, I. Savran, and Yu. V. Pershin,
                     Phys. Rev. B {\bf 83}, 193302 (2011).

\bibitem{Nunner11}  M. C. L\"{u}ffe, J. Kailasvuori, and T. S. Nunner,
arXiv:1103.0773.

\bibitem{Khodas04}  M. Khodas, A. Shekhter, and A. M. Finkelstein, Phys.
Rev. Lett. \textbf{92}, 086602 (2004).

\bibitem{Smirnov11}  S. Smirnov, Phys. Rev. B \textbf{83}, 081308 (2011).

\bibitem{Sinova04}  J. Sinova, D. Culcer, Q. Niu, N. A. Sinitsyn, T.
Jungwirth, and A. H. MacDonald, Phys. Rev. Lett. \textbf{92}, 126603 (2004); J.
Wunderlich, B. G. Park, A. C. Irvine, L. P. Zarbo, E. Rozkotova, P. Nemec,
V. Novak, J. Sinova, and T. Jungwirth, Science \textbf{330}, 1801 (2010).

\bibitem{disorder}  O. V. Dimitrova, Phys. Rev. B \textbf{71}, 245327 (2005); R.
Raimondi and P. Schwab, Phys. Rev. B \textbf{71}, 033311 (2005); A.
Khaetskii, Phys. Rev. Lett. \textbf{96} (2006) 056602; T. S. Nunner, N. A.
Sinitsyn, M. F. Borunda, V. K. Dugaev, A. A. Kovalev, Ar. Abanov, C. Timm, T.
Jungwirth, J.-I. Inoue, A. H. MacDonald, and J. Sinova, Phys. Rev. B \textbf{%
76}, 235312 (2007).

\bibitem{Ohno99}  Y. Ohno, R. Terauchi, T. Adachi, F. Matsukura, and H.
Ohno, Phys. Rev. Lett. \textbf{83}, 4196 (1999).

\bibitem{Dohrmann04}  S. Dohrmann, D. H\"{a}gele, J. Rudolph, M. Bichler, D.
Schuh, and M. Oestreich, Phys. Rev. Lett. \textbf{93}, 147405 (2004).
\bibitem{Hankiewicz06}  E. M. Hankiewicz, G. Vignale, and M. E. Flatt\'{e},
Phys. Rev. Lett. \textbf{97}, 266601 (2006).

\bibitem{Schreiber07}  L. Schreiber, D. Duda, B. Beschoten, G. G\"{u}ntherodt,
H.-P. Sch\"{o}nherr, and J. Herfort, Phys. Rev. B \textbf{75},
193304 (2007).

\bibitem{Belkov08}  V. V. Bel'kov, P. Olbrich, S. A. Tarasenko, D. Schuh, W.
Wegscheider, T. Korn, C. Schuller, D. Weiss, W. Prettl, and S. D. Ganichev,
Phys. Rev. Lett. \textbf{100}, 176806 (2008).

\bibitem{Crankshaw09}  S. Crankshaw, F. G. Sedgwick, M. Moewe, C.
Chang-Hasnain, H. Wang, and S.-L. Chuang, Phys. Rev. Lett. \textbf{102},
206604 (2009).

\bibitem{Tarasenko09}  S. A. Tarasenko, Phys. Rev. B \textbf{80}, 165317
(2009).

\bibitem{zhou09}  Y. Zhou and M. W. Wu, Solid State Communs. \textbf{49},
2078 (2009); Y. Zhou and M.W. Wu, EPL \textbf{89}, 57001 (2010).

\bibitem{Volkl} R. V\"{o}lkl, M. Griesbeck, S. A. Tarasenko, D. Schuh, W. Wegscheider, C. Schüller, and T. Korn,
                Phys. Rev. B {\bf 83}, 241306 (2011).

\bibitem{Hubner} J. H\"{u}bner, S. Kunz, S. Oertel, D. Schuh, M. Pochwala, H. T. Duc, J. F\"{o}rstner, T. Meier, and M. Oestreich,
                 Phys. Rev. B {\bf 84}, 041301 (2011).


\bibitem{Sih} V. Sih, R. C. Myers, Y. K. Kato, W. H. Lau, A. C. Gossard and D. D. Awschalom,
                 Nature Physics  {\bf 1}, 31 (2005).

\bibitem{dresselhaus55}  G. Dresselhaus, Phys. Rev. \textbf{100}, 580 (1955).

\bibitem{winkler04}  R. Winkler, \textit{Spin-Orbit Coupling Effects in
Two-dimensional Electron and Hole Systems} (Springer, New York, 2003).

\bibitem{Dyakonov86}  M. I. D'yakonov and V.Yu. Kachorovskii, Sov. Phys.
Semiconductors \textbf{20}, 110 (1986).

\bibitem{Muller08}  G. M. M\"{u}ller, M.L. R\"{o}mer, D. Schuh, W.
Wegscheider, J. H\"{u}bner, and M. Oestreich, Phys. Rev. Lett. \textbf{101},
206601 (2008); G. M. M\"{u}ller, M. Oestreich, M.L. R\"{o}mer, and J. H\"{u}%
bner, Physica E \textbf{43}, 569 (2010).


\bibitem{sherman03}  E. Ya. Sherman, Phys. Rev. B \textbf{67}, 161303(R) (2003).

\bibitem{sherman05}  E. Ya. Sherman, A. Najmaie, and J. E. Sipe, Appl. Phys. Lett.
\textbf{86}, 122103 (2005); H. Zhao, X. Pan, A. L. Smirl, R. D. R. Bhat, A.
Najmaie, J. E. Sipe, and H. M. van Driel, Phys. Rev. B \textbf{72}, 201302
(2005).

\bibitem{glazov10}  M. M. Glazov, E. Ya. Sherman, and V. K. Dugaev, Physica
E \textbf{42}, 2157 (2010).

\bibitem{Golub} L. E. Golub and E. L. Ivchenko, ArXiv:1107.1109.

\bibitem{Lyanda} Y. B. Lyanda-Geller, ArXiv:1107.3121.

\bibitem{Tang}  C. S. Tang, A. G. Mal'shukov, and K. A. Chao,
                Phys. Rev.  B {\bf 71}, 195314 (2005),
                C.-H. Lin, C.-S. Tang, and Y.-C. Chang,
                Phys. Rev. B {\bf 78}, 245312 (2008).

\bibitem{Zhang}  Q. Zhang, K. S. Chan, and Z. Lin,
          Appl. Phys. Lett. {\bf 98}, 032106 (2011).

\bibitem{Dugaev09}  V. K. Dugaev, E. Ya. Sherman, V. I. Ivanov, and J. Barna%
\'{s}, Phys. Rev. B \textbf{80}, 081301(R) (2009).

\bibitem{Rashba} E. I. Rashba, Phys. Rev. B {\bf 68}, 241315 (2003).

\bibitem{Tokatly}  I. V. Tokatly, Phys. Rev. Lett. \textbf{101}, 106601
(2008); I.~V.~Tokatly and E. Ya. Sherman, Ann. Physics \textbf{325}, 1104
(2010).

\bibitem{Pershin08}  This second-order process is qualitatively different
from the nonlinear photogalvanic effect in gyrotropic crystals (E. L.
Ivchenko and G. E. Pikus, JETP Lett. \textbf{27}, 604 (1978)) and from the
frequency doubling in the spin-Hall effect (Yu. V. Pershin and M. Di Ventra,
Phys. Rev. B \textbf{79}, 153307 (2009)).

\end{thebibliography}
\end{document}